# Temperature-controlled interlayer exchange coupling in strong/weak ferromagnetic multilayers: a thermo-magnetic *Curie-switch*

A. Kravets[1,2], A. N. Timoshevskii[2], B. Z. Yanchitsky[2], M. Bergmann[1], J. Buhler[1], S. Andersson[1], and V. Korenivski[1]

[1]*Nanostructure Physics, Royal Institute of Technology, 10691 Stockholm, Sweden*

[2]*Institute of Magnetism, National Academy of Science, 03142 Kiev, Ukraine*

**We investigate a novel type of interlayer exchange coupling based on driving a strong/weak/strong ferromagnetic tri-layer through the Curie point of the weakly ferromagnetic spacer, with the exchange coupling between the strongly ferromagnetic outer layers that can be switched, on and off, or varied continuously in magnitude by controlling the temperature of the material. We use Ni-Cu alloy of varied composition as the spacer material and model the effects of proximity-induced magnetism and the interlayer exchange coupling through the spacer from first principles, taking into account not only thermal spin-disorder but also the dependence of the atomic moment of Ni on the nearest-neighbor concentration of the non-magnetic Cu. We propose and demonstrate a gradient-composition spacer, with a lower Ni-concentration at the interfaces, for greatly improved effective-exchange uniformity and significantly improved thermo-magnetic switching in the structure. The reported magnetic multilayer materials can form the base for a variety of novel magnetic devices, such as sensors, oscillators, and memory elements based on thermo-magnetic *Curie-switching* in the device.**

Interlayer exchange coupling is one of the key fundamental characteristics of magnetic multilayers [1, 2], important for such large scale industrial applications as field sensors, magnetic recording, and magnetic random access memory [3-6]. In many cases it controls the magnetization switching in the system under the influence of external fields [7] or spin-polarized currents [8-10]. The oscillatory interlayer exchange coupling (RKKY) [11-13] is due to the conduction electrons mediating the spin transfer between the ferromagnetic layers, and is fixed in fabrication to be positive or negative in magnitude by selecting a suitable thickness of the nonmagnetic metal spacer. Once the coupling is set to antiparallel, an external switching field is necessary to change the state of the structure to parallel.

It is highly desirable to design multilayer materials where the interlayer exchange coupling is not fixed but rather controllable, on and off, by varying an external physical parameter, such as temperature. One such system is strong/weak/strong ferromagnetic sandwich (F/f/F) where the weakly ferromagnetic spacer (f) has a lower Curie temperature ($T_C$) than that of the strong ferromagnetic outer layers (F) [14,15]. Heating the structure through the $T_C$ of the spacer exchange-decouples the outer magnetic layers, so their parallel alignment below $T_C$ can be switched to antiparallel above $T_C$. This switching is fully reversible on cooling through the $T_C$, as the number of thermal magnons is reduced and the exchange spring in the spacer, aligning the outer F-layers, becomes stronger. This action can provide a spin switch or oscillator with intrinsic thermo-electronic control by the Joule heating of a transport current through the structure.

The key element in such an F/f/F sandwich is the weakly ferromagnetic spacer f, which should have a tunable in fabrication but well defined in operation $T_C$ and, preferably, a narrow ferromagnetic (f) to paramagnetic (P) transition. Diluted ferromagnetic alloys, such as Ni-Cu, with the $T_C$ in the bulk known to be easily tunable to near room temperature [16, 17] is the natural choice for the spacer material. However, the effects of thermal disorder on the magnetization and exchange coupling in thin-film multilayers are practically unexplored. In particular, the strong exchange at an F/P interface should be expected to suppress the thermal magnons in the spacer, driving the para-to-ferromagnetic transition due to the magnetic proximity effect [18,19], thereby resulting in a gradient of the effective magnetization and interatomic exchange in the spacer [20,21], as well as critically affecting the interlayer exchange coupling through the spacer. In this work we indeed find a pronounced ferromagnetic proximity effect at F/f(P) interfaces as well as propose and demonstrate experimentally a gradient spacer design (f*/f/f*), which significantly improves the thermo-magnetic switching behavior of the multilayer material and makes it attractive for technological applications.

Our choice for the diluted ferromagnetic alloy to be used as the spacer material is Ni-Cu. It is considered to be a well known system, at least in the bulk [16,17]. Our recent detailed studies of sputter-deposited Ni-Cu films confirmed the known general properties, but also revealed some pecular properties, such as exchange-induced phase separation at high Ni concentration $x$, above 70% of atomic Ni corresponding to the Curie temperature range above 100º C [22]. In this work, the concenration range of interest is $x$(Ni)≤70%, corresponding to the $T_C$ range of 100º C and below. In fact, the Ni-Cu concentrations with x<50% are non-magnetic at all temperatures in the bulk [16,17] or thick films [23]. The situation is quite different in thin film multilayers, as we show below.

Figure 1 shows the saturation magnetization of 60 nm thick $Ni_xCu_{100-x}$ films of varied composition, deposited at ambient temperature by dc magnetron sputtering on oxidized Si substrates, normalized to the saturation magnetization of pure Ni. The magnetization vanishes at room temperature for $x$≈62 at.%, where the Ni-Cu becomes paramagnetic in the bulk limit (here single-layer, 60 nm thick). Figure 1 also shows the saturation magnetization measured at 100º C, which vanishes at $x$≈70%. The ratio of the magnetization at these two temperatures, shown by the blue symbols in Fig. 1, has a sharp step at 70-74% Ni, indicating the optimum composition interval for exploiting the ferro-to-paramagnetic transition in the Ni-Cu alloy. We show, however, that this composition range must be significantly shifted to lower effective concentrations, if the Ni-Cu spacers are to be used for sharp on/off thermal switching in thin-film multilayers.

Perhaps the most informative way to investigate the properties of thin spacers as it relates to the interlayer exchange is to integrate them into a spin-valve type structure, AF/F/f/F, where one of the outer strongly ferromagnetic layers is pinned by an antiferromagnet (AF), and study the coupling/decoupling of the outer ferromagnetic layers (F) as a function of the spacer composition, thickness, and temperature. The method is not direct as to measuring the magnetization of the spacer, but is very sensitive and direct when it comes to the inter-layer exchange interaction of interest. Our material system is a Permalloy-clad spacer, with one of the Permalloy (Py) layers exchange-pinned to antiferromagnetic IrMn. Specifically, $Ir_{20}Mn_{80}$(12 nm)/$Co_{90}Fe_{10}$(2 nm)/$Ni_{80}Fe_{20}$(2 nm)/$Ni_xCu_{100-x}$(t nm)/$Ni_{80}Fe_{20}$(6 nm), henceforth $F_{pin}$/$Ni_xCu_{100-x}$(t nm)/F.

Figure 2 shows the key magnetic parameters of the $F_{pin}$/$Ni_xCu_{100-x}$(t nm)/F structures, measured at room temperature. Figure 2a shows magnetization loops for the $Ni_xCu_{100-x}$ spacer thickness of 6 nm and $x$ varied in the range from 0 to 72 at.%. For low Ni concentrations ($x$<35%), due to the absence of any significant magnetic coupling between the free and pinned F layers through the spacer, the magnetization loop consists of two well-separated transitions at approximately zero field and -480 Oe (Fig 2a, green), corresponding to switching of the free Py layer and the pinned ferromagnetic layer, respectively. With increasing Ni-concentration past $x$≈35%, the minor and major loops begin to merge (Fig. 2a, blue), indicating the onset of exchange coupling between the free and pinned outer layers. The middle points of the two magnetization transitions (minor and major) define the two exchange fields, $H_{ex1}$ and $H_{ex2}$, respectively. As the spacer becomes fully non-magnetic at low Ni-concentrations and does not mediate any exchange coupling, the unpinned Py layer becomes free to switch and $H_{ex1}$→0, while $H_{ex2}$ characterizes solely the strength of the AF-pining of the other ferromagnetic layer. Already at $x$≈52%, the two transitions merge significantly, indicating a substantial exchange coupling across the spacer. Interestingly, the $x$=52% composition for a single-layer Ni-Cu is non-magnetic (paramagnetic) at room temperature ($T_C$~10 K) and normally would not be expected to exchange couple the outer F-layers. These data indicate that a ferromagnetic order of significant strength is induced in the paramagnetic spacer on rather long length scales, several nanometers in this case. This induced ferromagnetism couples the outer layers, bringing together the two magnetic transitions, such that $H_{ex1}$ and $H_{ex2}$ merge. Thus, for this geometry, $H_{ex1}$ is a direct measure of the interlayer exchange coupling, while $H_{ex2}$ additionally reflects the strength of the antiferromagnetic pinning. For $x$>70%, a composition ferromagnetic at room temperature in the bulk, the minor and major loops merge into one (Fig. 2a, red). Fine-stepping through the low-concentration range, illustrated by the minor loops in Fig. 2b, shows that the onset of the interlayer exchange indeed is at $x \approx 35\%$, which is due to the vanishing Ni atomic magnetic moment in the Cu matrix, as detailed below.

The dependence of the two exchange fields, $H_{ex1}$ and $H_{ex2}$, on the Ni-concentration in a 6 nm thick Ni-Cu spacer is shown in Fig. 2c. The free and pinned ferromagnetic layers are fully decoupled up to $x=35\%$ (red symbols), at which point $H_{ex1}$ begins to increase in magnitude, first slightly and then substantially above $x=50\%$, even though the spacer is still intrinsically paramagnetic at this concentration. At $x \approx 70\%$ the two exchange fields merge into one, which is expected since the spacer is intrinsically ferromagnetic at 70% at RT.

Fig. 2d shows the thickness dependence of the exchange fields for a nominally (in the bulk) paramagnetic spacer composition of $x=56\%$. One can see that at 3 nm thickness the outer ferromagnetic layers are fully coupled and behave as one. For this composition, the interlayer exchange vanishes at approximately 9 nm in the spacer thickness. This is much greater that the interatomic spacing normally associated with direct exchange and indicates that the characteristic length scale for the induced ferromagnetic proximity effect under study is dictated by another mechanism, namely, thermally disordered lattice spins in the spacer by short-wave spin-waves on length scales of at least several lattice units.

In order to understand the mechanism involved as well as optimize the performance of the material we develop a full model of the F/f(P)/F multilayer from first principles, which takes into account the thermal spin disorder as well as the effect of Cu-dilution on the atomic magnetic moment of Ni in the performance-critical spacer layer.

Our model system is a Ni/Ni-Cu/Ni three-layer, in which the diluted magnetic alloy spacer is enclosed by bulk-like *fcc* Ni [001] (for making the calculations time efficient; qualitatively same behavior is obtained with Permalloy outer layers). Both Ni and Cu atoms in the three-layer structure occupy the sites of the *fcc* lattice, and are distributed randomly within each monolayer in the spacer. The number of atomic monolayers in the spacer is denoted by $N_f$. The atomic concentration of $N_i$ in $i$-th monolayer is denoted by $c_i$. The Ni atoms interact magnetically by the standard isotropic Heisenberg interaction. Cu-Cu and Ni-Cu exchange interactions are neglected since the magnetic moment of Cu is negligible (Cu does not polarize in Ni). The local atomic magnetic moment of Ni, $m_{loc}(z)$, is a function of the number of the nearest neighbor Ni atoms, $z$. For obtaining the effective (measurable) magnetic characteristics of the structure we use the mean field model and take the average Ni-magnetic moment to be the same within one monolayer, $m_i=m(z_i)$. The effective magnetic field is [24]:

$$H_i = -\sum_j J_j n_{i+j} c_{i+j} m_{i+j},$$

where the sum is over monolayers, $J_i$ – the Ni-Ni exchange interaction, and $n_i$ – the coordination number. The magnetization at a given temperature $T$ is given by [24]: $m_i=L(m_i H_i/k_B T)$, where $L(x)$ is the Langevin function and $k_B$ – the Boltzmann constant.

The unknown exchange integrals $J_i$ were obtained for 2 coordination spheres of the *fcc* lattice. In this the total energies of 3 superstructures of *fcc* Ni were calculated for the ferro- (F), antiferro- (AF), and antiferro-double (AFD) [25] types of magnetic ordering.

With the magnetic energy in the Heisenberg form and assuming the magnetic moment of Ni independent of its direction, the following expressions for the exchange interaction are obtained: $J_1=(E_F-E_{AF})/8$; $J_2=(E_F+E_{AF}-2E_{AFD})/4$, where $E_F$, $E_{AF}$, and $E_{AFD}$ are the full energies of the superstructures.

The total energies of the structures were obtained using DFT approach and the Wien2k FLAPW code [26]. The GGA exchange-correlation potential was the same as in Ref. 27. The radius of the MT-spheres was 2.2 atomic length units. The electron density was computed for 63 $k$-points in the irreducible parts of the first Brillouin zone. The obtained exchange integrals were: $J_1$=-6.15 meV and $J_2$=-17.01 meV.

For obtaining the $m_{loc}(z)$ dependence, the electronic structure of three special quasi-random superstructures [28], which model random bulk Ni-Cu alloy, were calculated. The stoichiometries of the structures were: $Ni_{25}Cu_{75}$, $Ni_{50}Cu_{50}$, and $Ni_{75}Cu_{25}$. Figure 3a (solid circles) shows the values of $m_{loc}$ calculated by the FLAPW method. For calibration purposes, the slope of the Curie temperature of bulk Ni-Cu alloy was calculated and agreed well with the experiment [17], as shown in the inset to Fig. 3a. The interesting result in the obtained $m_{loc}(z)$ is that Ni becomes essentially nonmagnetic in the Ni-Cu alloy at concentration of approximately 30%-Ni. This has important implications for optimizing the spacer material, as discussed below.

The key for efficient operation of a spin-thermo-electronic F/f/F valve is a small width of its Curie transition. The green line in Fig. 3b shows the calculated magnetization per Ni atom of bulk $Ni_{80}Cu_{20}$ alloy. The blue curve shows the temperature dependence of the magnetization of a homogeneous Ni-Cu spacer with Ni outer layers. The spacer consists of 38 monolayers (approx. 7 nm) and is placed between outer planes of *fcc* Ni. At T=0K the magnetization is equal to the local moment of the Ni atom $m_{loc}$ for this composition (Fig. 3a). When the spacer is enclosed by strongly ferromagnetic outer electrodes (Ni), the spacer's ferromagnetic state greatly extends in temperature, vanishing completely only above 0.9 $T_C$ (Ni) (Fig. 3b, blue). This means that the outer electrodes are strongly exchange-coupled at the nominal $T_C$ of the spacer alloy (0.58), marked as the inflection point, $M_t$ ($T_t$). The effective transition extends over a broad interval of 0.3-0.4 $T_C$(Ni) above the effective transition point of 0.58. The reason for this extended transition is the strong ferromagnetic order induced in the spacer in proximity to the interfaces, as shown by the open symbols in Fig. 3c, for two characteristic temperatures. At $T_t$ the moment at the interface is enhanced 4-fold compared to that in the center of the spacer, with a strong variation in the effective $T_C$ across the thickness. The proximity length is an order of magnitude greater than the atomic spacing, so the induced magnetization penetrates through all of the spacer thickness. The result is non-zero magnetic exchange between the outer ferromagnetic layers, well above the intrinsic Curie point of the spacer material. This proximity effect should be universal for F/f/F tri-layers and sets a fundamental limitation on the width of the Curie transition of the weak ferromagnet incorporated in the multilayer.

It is highly desirable for device applications to narrow the magnetic transition in the spacer. Using the above detailed understanding of the highly non-uniform magnetization profile at the F/f interface, we have designed a *gradient-spacer*, in which the magnetic-atom concentration is reduced at the interfaces. This efficiently suppresses the proximity effect and makes the magnetization distribution much more uniform, as shown in Fig. 3c with solid symbols for a gradient spacer with the interface Ni concentration reduced from 80% to 65%.

This change in the spacer layout has a dramatic effect on the simulated transition width, as shown in Fig. 3b (red). The magnetization at the inflection point is 5 times smaller for the uniform-spacer design, which translates into an order of magnitude sharper Curie transition for the tri-layer, comparable in width with that in the ideal spacer (uniform, bulk-like; green in Fig. 3b).

In order to experimentally demonstrate the gradient-spacer effect proposed above, we have fabricated a range of valves, in which the spacer itself had a tri-layer structure, f*/f/f*, with the buffer layers f* of different thickness and Ni-content compared to the inner spacer layer f. The inner layer f had a fixed thickness and concentration of 6 nm and 72%, respectively. This new layout is illustrated in the inset to Fig. 4, which shows the phase map of the resulting proximity effect. The vertical scale gives the thickness of the buffer layer f* for a given concentration and temperature, at which the outer Py layers fully decouple, determined in the same fashion as in Figs. 2. The phase map thus gives the operating area for a Curie-valve based of the gradient-spacer design.

Interestingly, the scaling is logarithmic and shows that the thinnest layers decouple only below $x$=30%, where the Ni atoms become, in fact, nonmagnetic. We believe that this finding has high relevance for the RKKY interaction in this system in the thin-spacer limit studied previously [29, 30]. The RKKY interlayer coupling through thin Cu spacers was interpreted to withstand paramagnetic Ni impurities up to approximately 35% Ni, vanishing at higher concentrations. We suggest that the mechanism behind the strong RKKY and its subsequent vanishing at higher Ni-content was instead the loss of the atomic moment on Ni below 30% Ni in Cu, detailed in our simulation results above (Fig. 3a).

Having established the key physical parameters of the new gradient spacer design, below we demonstrate its greatly improved thermo-electronic characteristics. Figure 5 compares the temperature dependence of the magnetization of the two spacer layouts, with uniform and gradient-type composition. The samples were heated to 100 ºC (to just above the bulk- $T_C$ of the inner spacer material, but below any significant reduction in the AF pinning), after which a reversing field of 50 Oe was applied in order to switch the free Py layer, and the temperature was gradually decreased to room temperature while the magnetization was recorded. As a result, the spacer acts as an exchange spring of increasing strength, which rotates the free Py layer during the cooling.

The Curie transition (para-to-ferromagnetic) is very broad in the uniform-spacer multilayer. In fact, the rotation of the free layer is far from complete at 100º C, even for the relatively thick spacer (20 nm), due to the residual proximity-induced interlayer exchange. In stark contrast, the gradient-spacer sample fully exchange-decouples into the antiparallel state of the outer Py layers at 90 ºC (the Curie point of the inner spacer material with $x=72\%$), and has a sharp transition into the parallel state of the multilayer at RT. The 20-80% width of the transition is approximately 20 degrees, same as the full-width-at-half-maximum width, and several times narrower than that for the uniform spacer. This result is in good agreement with the theoretically predicted behavior.

It is informative to note that the thermo-magnetic switching demonstrated herein can have significant advantages over the recently developed and very promising thermally-assisted switching, used in the memory technology, based on thermally controlling antiferromagnetic exchange pinning [31]. One advantage is the the Curie point of a diluted ferromagnet can be easily varied in the desired range and is not fixed to the Neel (or blocking) temperature of the antiferromagnet. Furthermore, the ferro-to-paramagnetic transition typically is fully reversible, does not involve spin "blocking", and therefore should not suffer from training effects present at the exchange-biased F/AF interface.

In conclusion, we have investigated strong/weak/strong ferromagnetic tri-layers where the interlayer exchange coupling is controlled by driving the material through the Curie point of the spacer. The resulting exchange coupling between the strongly ferromagnetic outer layers can be switched on and off, or varied continuously in magnitude by controlling the temperature of the material. This effect is explained theoretically as due to induced ferromagnetism at F/f(P) interfaces. It is shown that the atomic magnetic moment and the effective interatomic exchange coupling are highly non-uniform throughout the spacer thickness, especially in the proximity to the strongly ferromagnetic interfaces. This critically affects the interlayer exchange coupling and the ability to control it thermo-electronically. We have proposed and demonstrated a new, gradient-type spacer, having a significantly narrower Curie transition and distinct thermo-magnetic switching. The demonstrated multilayer material can form the base for a variety of novel magnetic devices based on spin-thermo-electronic switching.

We gratefully acknowledge financial support from EU-FP7-FET-Open through project Spin-Thermo-Electronics.

**Figure captions:**
1. Saturation magnetization of thick $Ni_xCu_{100-x}$ alloy films normalized to that of pure Ni, as a function of Ni content, measured at room temperature (RT) and 100º C. The solid lines are guides to the eye. Blue symbols show the ratio of the magnetization at the two temperatures, which becomes zero at the boundary of the thermo-magnetic operating region.

2. (a) Magnetization loops of $F_{pin}/Ni_xCu_{100-x}$(6 nm)/F tri-layers having 35, 52 and 70 Ni at.% content. The exchange fields, $H_{ex1}$ and $H_{ex2}$, for the free and pinned ferromagnetic layers are defined as the midpoints of the respective transitions. (b) Magnetization minor loops of $F_{pin}/Ni_xCu_{100-x}$(6 nm)/F tri-layers with different Ni-Cu spacer compositions. Exchange fields of the two outer ferromagnetic layers versus (c) nickel concentration in a 6 nm thick Ni-Cu spacer and (d) thickness of the Ni-Cu spacer for $x$=56 at.%. The solid lines are guides to the eye.

3. (a) Calculated local atomic magnetic moment of Ni atoms in $Ni_xCu_{100-x}$ alloy as a function of the number of the Ni atoms in the first coordination sphere, normalized to the moment of bulk *fcc* Ni of $m_0(Ni)=0.63\mu_B$. The solid lines are guides to the eye. Inset shows the calculated and experimental [17] slopes of $T_C$ in the bulk. (b) Calculated magnetization per nickel atom versus temperature for a bulk $Ni_{80}Cu_{20}$ alloy (green), uniform $Ni_{80}Cu_{20}$(7 nm) spacer (blue) and a gradient composition spacer $Ni_{65}Cu_{35}$(1 nm)/$Ni_{84}Cu_{16}$(5 nm)/ $Ni_{65}Cu_{35}$(1 nm) (red), the spacers enclosed by outer Ni layers. The inflection points, where the transition is steepest as defined by the second derivative changing sign, are marked with $M_t(T_t)$. (c) Calculated magnetization profiles in a uniform $Ni_{80}Cu_{20}$(38 ML) (open symbols) and gradient $Ni_{65}Cu_{35}$(4 ML)/ $Ni_{84}Cu_{16}$(30 ML)/ $Ni_{65}Cu_{35}$(4 ML) (solid symbols) composition spacers for two different temperatures near the respective Curie points.

4. Thickness of the interfacial buffer layer $f^*$ of the weakly ferromagnetic spacer at which the outer Py layers decouple, as a function of its Ni-content, for room temperature and 100º C. The thickness and Ni-concentration of the inner spacer layer f are 6 nm and 72 at.%, respectively.

5. Magnetization of two uniform $Ni_{72}Cu_{28}$ spacer multilayers, with 10 nm (us1) and 20 nm (us2) thick spacers, and of a gradient $Ni_{50}Cu_{50}$(4 nm)/$Ni_{72}Cu_{28}$(6 nm)/$Ni_{50}Cu_{50}$(4 nm) spacer multilayer (gs), normalized for a direct comparison. The solid lines are guides to the eye. The arrows in the illustrations of the right panel show the respective relative orientations of the magnetic moments in the center of the individual layers of the exchange coupled F/f/F multilayer.

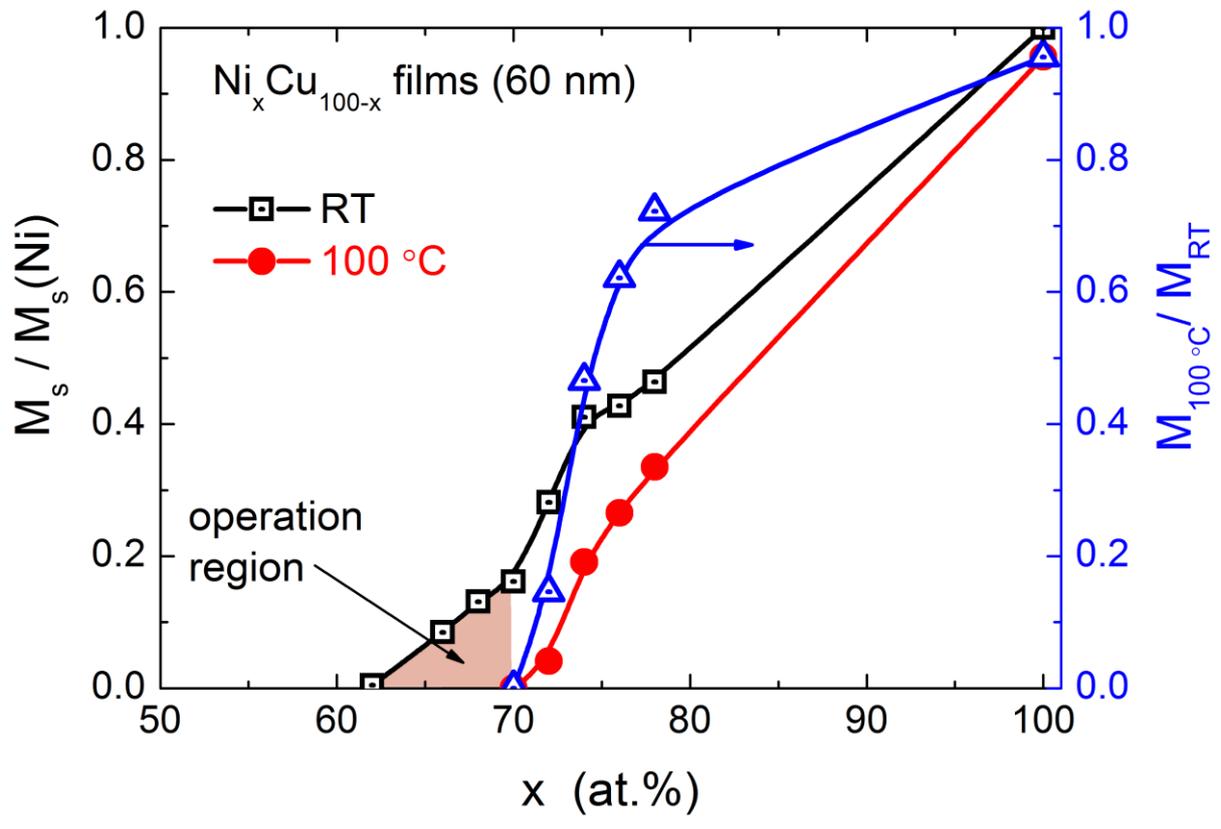

FIG. 1

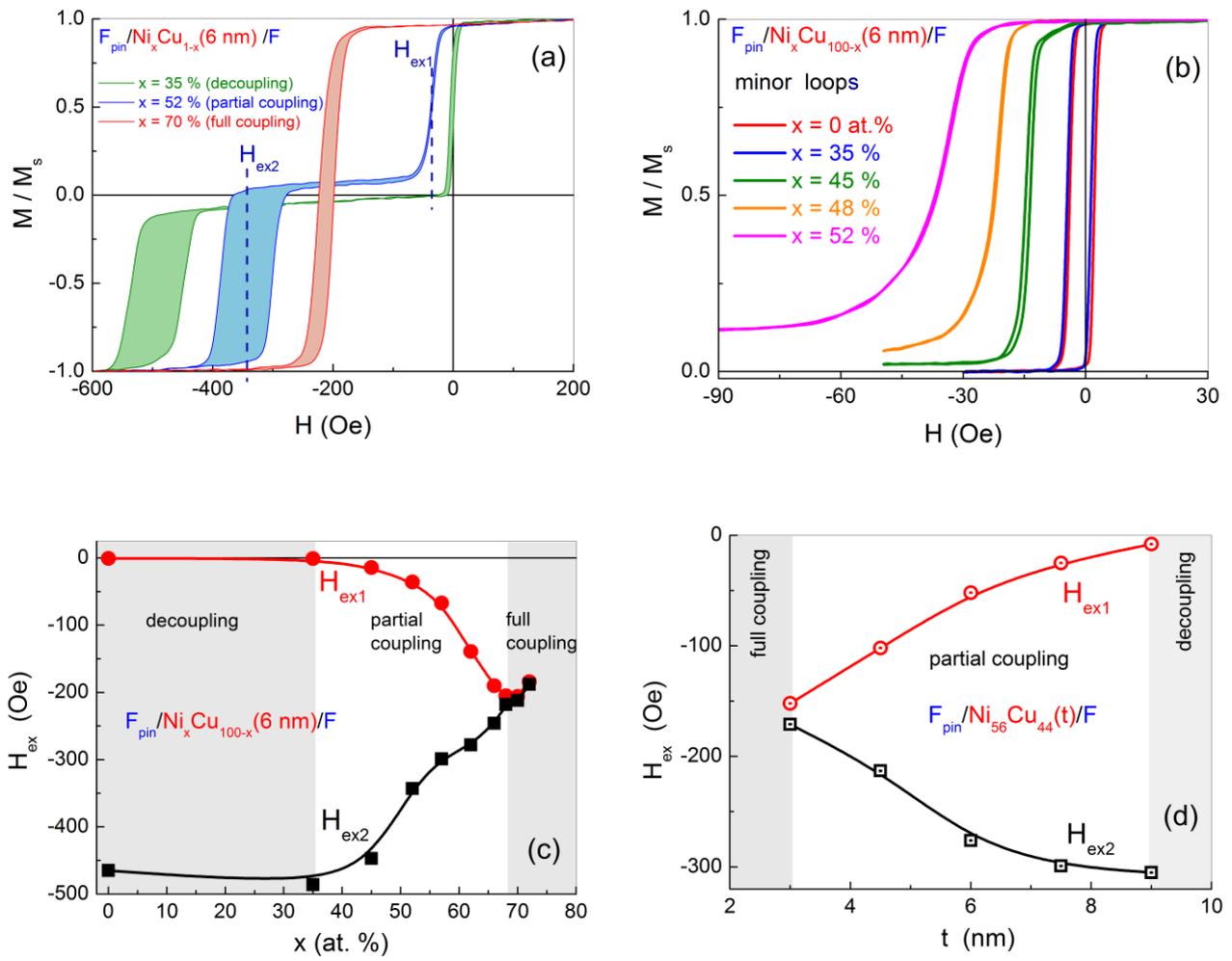

FIG 2.

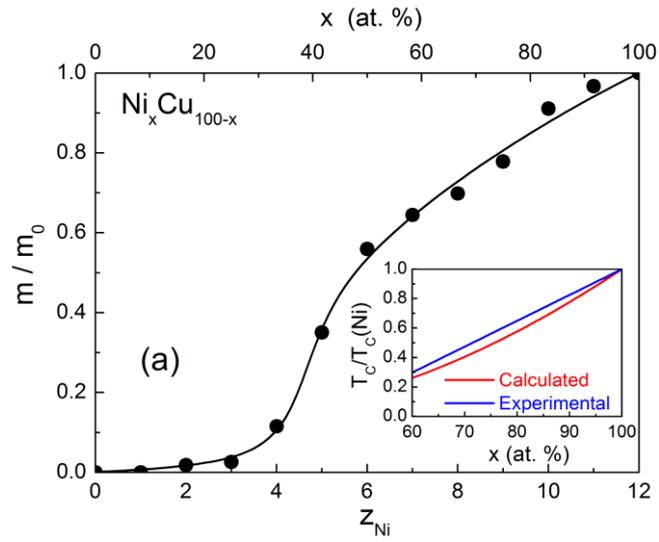

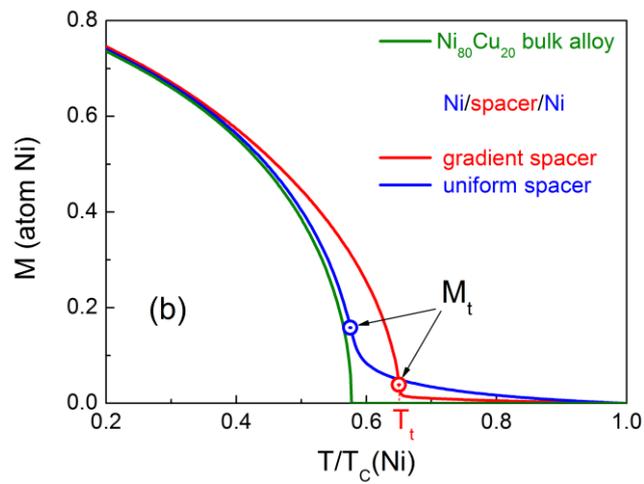

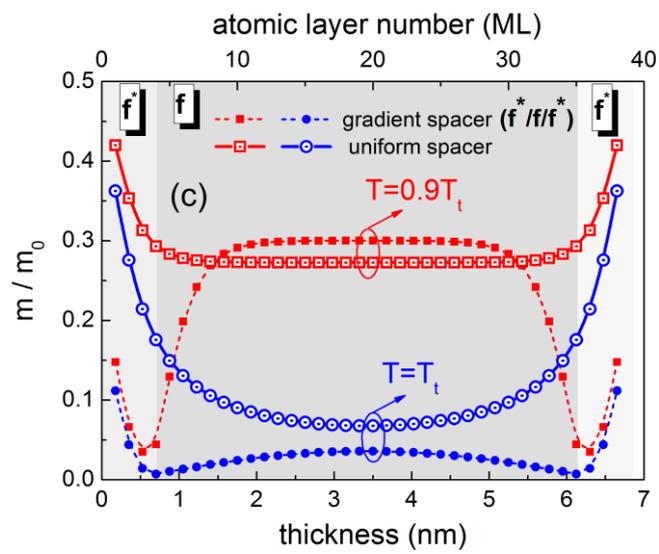

FIG. 3

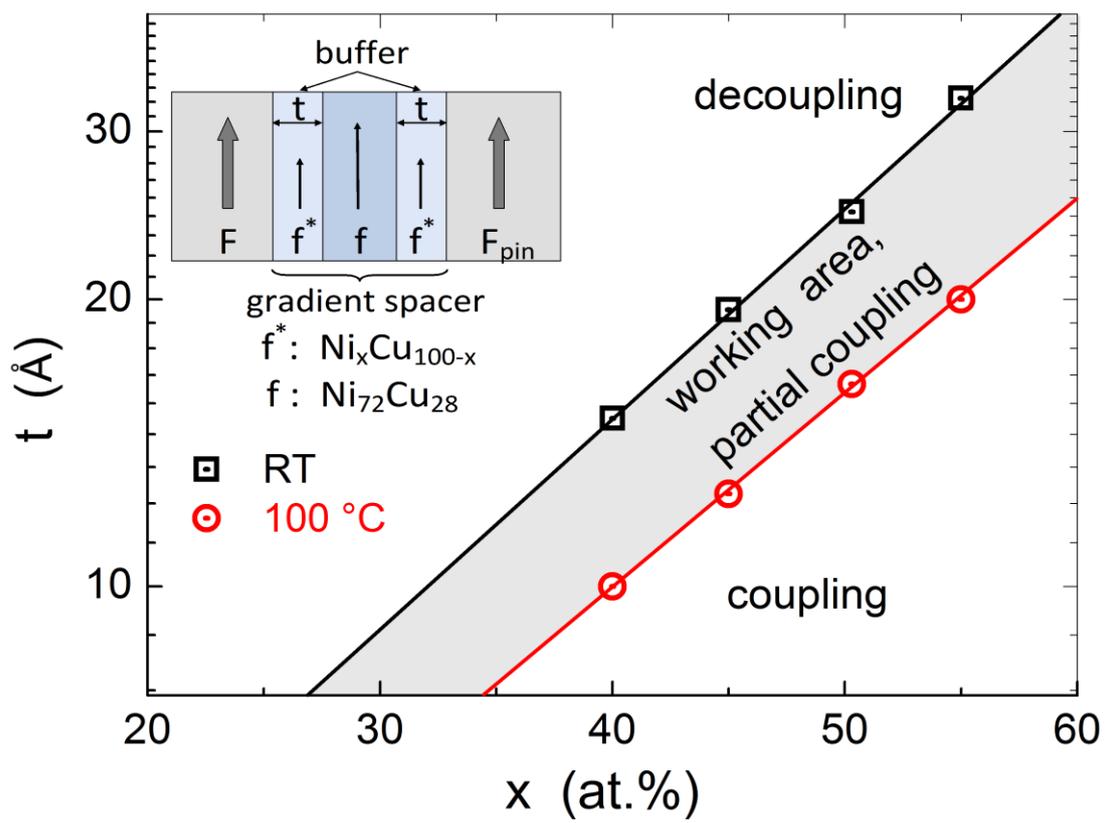

FIG. 4

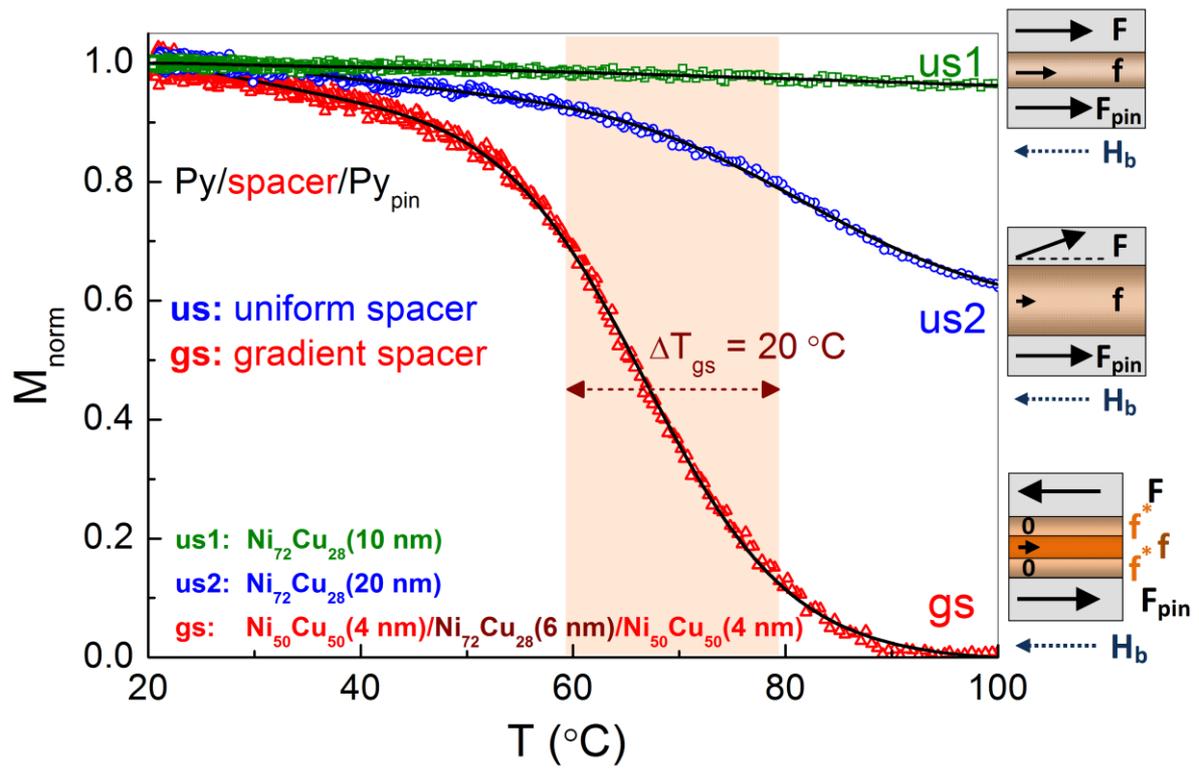

FIG. 5